\documentclass{article}
\usepackage{graphicx}
\usepackage{slashed}
\usepackage[hypertex]{hyperref}  
\usepackage{amssymb}

\newcommand{\Mvariable}[1]{}

 \providecommand{\imag}[1]{\,i\,}
 
\usepackage{color}

\author{Ivo de Medeiros Varzielas $^{(a)}$ \thanks{ivo@cftp.ist.utl.pt}, Graham G. Ross $^{(b)}$\thanks{g.ross@physics.ox.ac.uk}, Mario Serna $^{(b)}$\thanks{serna@physics.ox.ac.uk} \\
\small
(a) CFTP, Departamento de F\'{i}sica, Instituto Superior T\'{e}cnico,
Av. Rovisco Pais, 1, 1049-001 Lisboa, Portugal \\
\small (b) Rudolf Peierls Centre for Theoretical Physics, University of Oxford,
1 Keble Road, Oxford, OX1 3NP, U.K.
\normalsize
}

\title{Quasi-degenerate neutrinos and tri-bi-maximal mixing}

\begin{document}

\maketitle

\begin{abstract}
We consider how, for quasi-degenerate neutrinos with tri-bi-maximal mixing
at a high-energy scale, the mixing angles are affected by radiative running
from high to low-energy scales in a supersymmetric theory. The limits on the
high-energy scale that follow from consistency with the observed mixing are
determined. We construct a model in which a non-Abelian discrete family
symmetry leads both to a quasi-degenerate neutrino mass spectrum and to near
tri-bi-maximal mixing.
\end{abstract} 

\section{Introduction}

Neutrino oscillation data is at present consistent \cite{Maltoni:2004ei,
Abe:2008ee} with just three light neutrinos with near tri-bi-maximal (TBM)
mixing between flavours \cite{Wolfenstein:1978uw, Harrison:2002er,
Harrison:2002kp, Harrison:2003aw, Low:2003dz}. However the nature of the
mass spectrum is still not established, being consistent with either a
normal or an inverted hierarchy. Moreover, although the magnitude of the
mass squared difference between neutrinos is reasonably well determined, the
absolute scale of mass is not, being consistent with both a strongly
hierarchical spectrum or a quasi-degenerate (QD) spectrum.

Radiative running is especially important for QD neutrinos, as the effects
on mixing angles are larger for QD neutrinos than in the hierarchical case.
This was stressed in \cite{Ellis:1999my, Casas:1999tp} where the mixing
favoured at the time, bi-maximal mixing, was studied in-depth. More recent
studies of mixing angles running include \cite{Antusch:2003kp,
Plentinger:2005kx, Dighe:2006sr, Boudjemaa:2008jf} (and references
therein). Here we discuss radiative corrections to TBM mixing, assuming that
it arises through new physics, such as a family symmetry, at a high-energy scale.
We determine how high, in a supersymmetric extension of the
Standard Model, the initial energy scale can be while maintaining near TBM
mixing at the low-energy scales relevant to oscillation experiments. The
main difference from existing work is that emphasis is placed on the energy
scales rather than on the resulting low-energy angles. Specifically, we set
the angles to their TBM values at high-energy scales, run the angles to
low-energy and iterate the process to find the highest-energy scale that
still keeps the low-energy angles within current experimental bounds. The
process is then repeated for different points of the parameter space, and
the results are presented as a contour plot in the $m_{\nu _{i}}-\tan \beta $
plane ($i=1$ for normal and $i=3$ for inverted hierarchy).

The underlying question raised by the observed near TBM mixing is
the origin of the pattern and the reason it is so different from quark
mixing. Models based on family symmetries, particularly discrete non-Abelian
family symmetries, have been constructed to explain this pattern, e.g. \cite{deMedeirosVarzielas:2006fc, King:2006np}. In
these models the difference between the quark and lepton sector follows
naturally from the see-saw mechanism together with a strongly hierarchical
right-handed neutrino Majorana mass spectrum. However these models only
apply to the case of an hierarchical neutrino mass spectrum. Here we discuss
how a discrete non-Abelian family symmetry can also give rise to near TBM
mixing for the case of a QD spectrum.

\section{Radiative corrections to TBM mixing}

Family-symmetry models are typically constructed at some high scale, $M_{F}$,
at which the model specifies relationships among parameters. To compare
the predictions to low-energy data, radiative effects should be considered
through the use of the renormalization group equations. When there is a
strong hierarchy, it is often the case that these running effects do not
change the mixing angles by much \cite%
{Antusch:2003kp,Plentinger:2005kx,Dighe:2006sr, Boudjemaa:2008jf}. In the
case of QD neutrinos, however, the mixing angles can change a lot with the
energy scale, to the point of erasing any special structure arranged by a
family symmetry. For model-building purposes it is very important to know
the highest-energy scale at which we can start with TBM mixing and still be
consistent with mixing-angle data after running the angles down to the
low-energy scale $M_{Z}$ (the Z-boson mass scale).

The Standard Model (SM) suffers from the hierarchy
problem associated with the need to keep electroweak breaking much below
the Planck scale. This problem is evaded if the theory is supersymmetric, with
supersymmetry broken close to the electroweak scale. For this reason we
consider the radiative corrections to neutrino masses and mixing in the
context of the minimal supersymmetric extension of the Standard Model
(MSSM). We specify the low-energy boundary conditions of the
renormalization group equations to be consistent with the three gauge
coupling constants and the quark and lepton masses \cite{PDBook2008}.
We assume an effective SUSY scale of $M_{S}=500$ GeV. We use the SM
renormalization group equations below $M_{S}$ and the MSSM renormalization
group equations above $M_{S}$. The only boundary condition set at the
family-symmetry breaking scale $M_{F}$ is exact TBM mixing for the leptons \footnote{We ignore the small departures from TBM at the high scale which may arise from diagonalising the charged-lepton mass matrix \cite{Plentinger:2005kx, Antusch:2005kw}.}.
The neutrino masses are set at the low-energy boundary relative to
the lightest neutrino mass state ($m_{\nu_{1}}$ with a normal hierarchy and $m_{\nu_{3}}$ with an inverted hierarchy).
We keep $\left| \Delta m^2_{12} \right|$ the solar mass difference and $\left| \Delta m^2_{23} \right|$ the atmospheric mass
difference.
\begin{figure}[tbp]
\centerline{\includegraphics[width=3.3in]{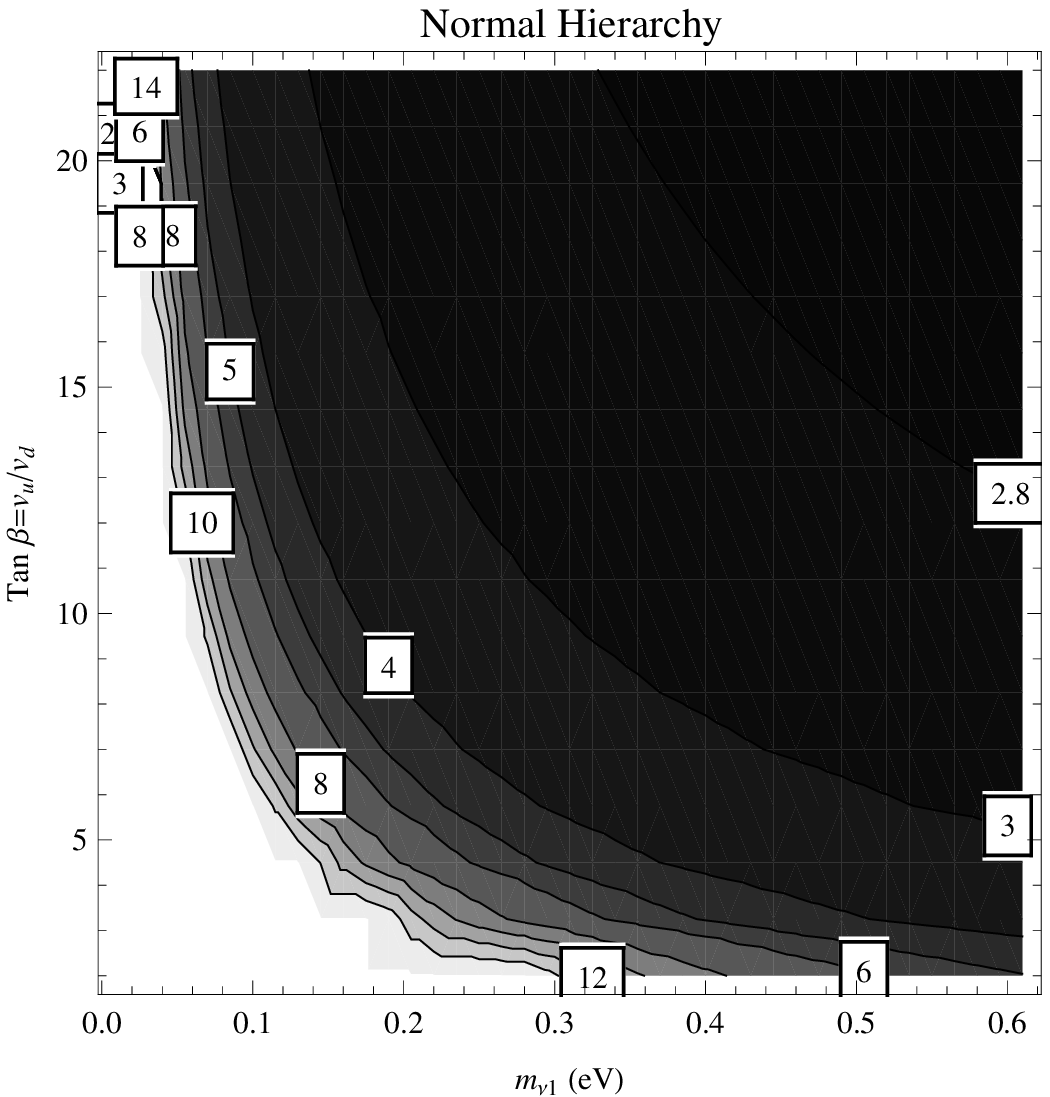}\ \ \includegraphics[width=3.3in]{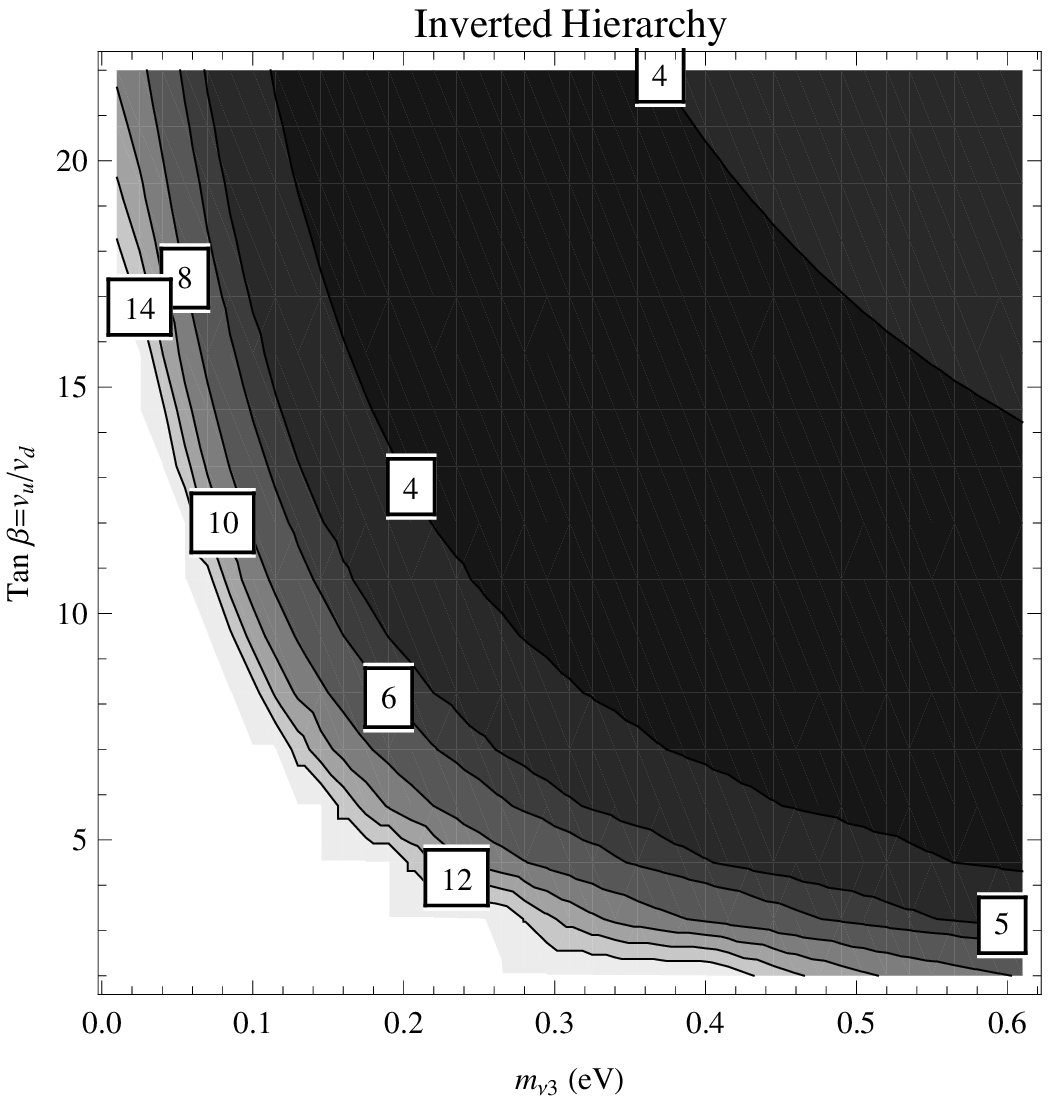}}
\caption{Shows contours of $\mathrm{Log}_{10}(M_{F})$ where $M_{F}$ is the
highest-energy family-symmetry breaking scale at which we can set TBM and
have the neutrino mixing within $4\protect\sigma $ of the low-energy
observed values. The white region in the lower left of the contour plots are the regions where $M_F$ can be greater than $10^{16}$ GeV.}
\label{FigContourSUSY}
\end{figure}

Figure \ref{FigContourSUSY} shows two contour plots.
For the normal hierarchy the plot shows $m_{\nu_1}$ versus $\tan
\beta$ and for the inverted hierarchy shows $m_{\nu_3}$ versus $\tan \beta$.
The contours specify $\mathrm{Log}_{10} M_F$ where $M_F$ is the
highest-energy family-symmetry breaking scale at which we can set TBM mixing
and have the low-energy mixing angles consistent to within $4 \sigma$ of the
low-energy observations. The solar mixing angle $\theta_{12}$ is the most
sensitive to radiative corrections. Exact TBM mixing gives $\tan^2
\theta_{12}=0.5$, and our $4 \sigma$ requirement at low energy translates to
$\tan^2 \theta_{12} = 0.47 \pm 0.2$ \cite{Abe:2008ee}.

The difference between the two graphs can mostly be understood by the slight bias of the observational data ($\tan^2\theta_{12} < 0.5$) and the opposite directions in which the $\tan^2\theta_{12}$ runs between the normal and inverted hierarchies.
Starting with perfect TBM mixing at the scale $M_F$ a normal hierarchy has $\tan^2\theta_{12}$  become larger as the the renormalization scale becomes smaller.
Once one falls below $M_S$, then $\tan^2\theta_{12}$ begins to get smaller as the renormalization scale goes down the during the final leg.  The inverted hierarchy has the opposite behavior. Because there is a longer region of supersymmetric running, there is more parameter space of $m_{\nu_3}$ versus $\tan \beta$ compatible with $M_F \ge 10^{16}$ GeV. The slight bulge visible in the upper-right of the inverted-hierarchy contour plot with $M_F \approx 10^4$ GeV
is due to the opposite directions of the supersymmetric running and the standard model running on $\tan^2 \theta_{12}$ in the region where $M_F \approx M_S$.

The contours in Figure \ref{FigContourSUSY} hold implications for QD TBM
family-symmetry models. For $m_{\nu _{1}}>0.1$ eV, the neutrino spectrum is
referred to as quasi-degenerate (QD) \cite{Vogel:2006sq} \footnote{%
We define QD as $m_{\nu _{1}}>0.1$ eV because above this value the $\beta
\beta _{0\nu }$ constraints for differing hierarchies and phases converge to
a common region, as shown in figure \ref{doublebeta}.}. If cosmological observations are considered, they
constrain the sum of the neutrinos $\sum_{i}m_{\nu _{i}}\leq 0.42$ eV at the
$95\%$ confidence level \cite{Tegmark:2005cy}. This implies $m_{\nu
_{1}}\leq 0.14$ eV which excludes the right half of Figure \ref%
{FigContourSUSY}. The remaining allowed narrow strip is consistent with the non-observation
of neutrinoless double beta decay $\beta \beta _{0\nu }$ which places a
limit of $m_{ee}<0.34$ eV. Uncertainties in nuclear matrix element weaken
this bound by about a factor of $3$. If we believe that the family-symmetry
scale is greater than $M_{F}>10^{10}$ GeV, and hypothesize a model which
leads to a QD neutrino spectrum with normal hierarchy, then $\tan \beta <6$ (or $\tan \beta < 8$ for a model with inverted hierarchy). In contrast, if a
normal hierarchy model has $\tan \beta >6$  (or $\tan \beta > 8$ for an inverted hierarchy model), then the lightest neutrino need be less than $0.1$ eV and therefore hierarchical.

\section{A discrete non-Abelian family symmetry model of QD neutrinos with
TBM mixing}

As stressed in \cite{Barbieri:1999km} an underlying $SO(3)$ family symmetry
readily leads to a near degenerate neutrino mass spectrum. In their model
the chiral superfields, $L^{i}$ (where $i$ is the $SO(3)$ family index),
contain the lepton doublets and transform as triplets under the $SO(3)$
group. The chiral superfields
containing the conjugates of the right-handed electron, muon and tau, 
respectively $e^{c}$, $\mu^{c}$ and $\tau^{c}$, are $SO(3)$ singlets. The effective Majorana neutrino mass is
constrained by the symmetry and comes from the superpotential 
\begin{equation}
W_{eff}=y_{0}(L^{i}L^{i})H_{u}H_{u}/M  \label{so3}
\end{equation}%
where $H_{u}$ is the supermultiplet containing the Higgs field whose vacuum
expectation value (VEV), $\left\langle H_{u}\right\rangle =v$, is
responsible for up quark masses in the MSSM and $M$ is the messenger scale
associated with the mechanism generating this dimension 5 term (in the Type
II see-saw it is the mass of the exchanged isotriplet Higgs field).

The important point to be taken from eq.(\ref{so3}) is that the family
symmetry forces the three light neutrinos to be degenerate. Small departures
from degeneracy result when the $SO(3)$ family symmetry is broken. In what
follows we will show how this can naturally lead to a mass mixing matrix
which gives near TBM mixing. This is done through the breaking of the family
symmetry by the non-vanishing vacuum expectation values (VEVs) of familon
fields, denoted as $\phi _{A}^{i}$, where the $A=3$, $23$, $123$ labels three
distinct fields and serves as a reminder of their VEV directions which are
given by 
\begin{equation}
\left\langle \phi _{3}\right\rangle =\left( 
\begin{array}{c}
0 \\ 
0 \\ 
a%
\end{array}%
\right) \ \left\langle \phi _{23}\right\rangle =\left( 
\begin{array}{c}
0 \\ 
-b \\ 
b%
\end{array}%
\right) \ \ \ \left\langle \phi _{123}\right\rangle =\left( 
\begin{array}{c}
c \\ 
c \\ 
c%
\end{array}%
\right)  \label{eq:P123 vev}
\end{equation}%
where $a,b$ and $c$ are complex parameters. Table \ref{ta:SO3} lists the
full set of supermultiplets and their symmetry properties under the $SO(3)$
symmetry extended by a further set of symmetries $G=Z_{3R}\times Z_{2}\times
U_{\tau }(1)$ which limit the terms that can appear in the superpotential.
$Z_{3R}$ is a discrete $R-$symmetry which ensures the familon fields
are moduli and cannot appear in the superpotential except coupled to
``matter'' fields carrying non-zero $R-$charge. The $U_{\tau }(1)$ symmetry is
introduced to distinguish the third family of leptons from the first two. In
practice it also explains why the mixing in the charged-lepton sector is
different from that in the neutrino sector which leads to near
tri-bi-maximal mixing.

\begin{table}[tbp]
\begin{center}
\begin{tabular}{|c|c|c|c|c|}
\hline
Field & $SO(3)$ & $Z_{3R}$ & $U_{\tau }(1)$ & $Z_{2}$ \\ \hline
$L^{i}$ & 3 & 1 & 0 & + \\ 
$e^{c}$ & 1 & 1 & 0 & + \\ 
$\mu ^{c}$ & 1 & 1 & 0 & - \\ 
$\tau ^{c}$ & 1 & 1 & -1 & + \\ 
$H_{u,d}$ & 1 & 0 & 0 & + \\ \hline
$\phi _{3}^{i}$ & 3 & 0 & 1 & + \\ 
$\phi _{23}^{i}$ & 3 & 0 & 0 & - \\ 
$\phi _{123}^{i}$ & 3 & 0 & 0 & + \\ \hline
$X$ & 1 & 2 & 0 & - \\ \hline
\end{tabular}%
\end{center}
\caption{Assignment of the fields under the $SO(3)$ family symmetry.}
\label{ta:SO3}
\end{table}
The special structure of the VEVs in eq(\ref{eq:P123 vev}) is what will
generate TBM mixing and is clearly the most important aspect of the model.
This can happen naturally if the underlying family symmetry is not $SO(3)$
but a discrete non-Abelian subgroup. We will discuss below the nature of
this symmetry and the vacuum alignment leading to eq(\ref{eq:P123 vev}) (the 
$X$ field of Table \ref{ta:SO3} is introduced to facilitate this vacuum
alignment), but first we show that it does generate approximate TBM mixing.

The leading terms in the superpotential responsible for neutrino masses that
are invariant under the family symmetries are given by 
\begin{equation}
W_{\nu }=y_{0}(L^{i}L^{i})H_{u}H_{u}+y_{\odot }(\phi
_{123}^{i}L^{i})^{2}H_{u}H_{u}+y_{@}(\phi _{23}^{i}L^{i})^{2}H_{u}H_{u}.
\label{win}
\end{equation}%
where we have suppressed the messenger scale. Note that due to the $Z_{2}$
factor there are no cross terms involving $\phi _{23}\phi _{123}$ \cite%
{Ross:2007zz, deMedeirosVarzielas:2008en} and due to the $U_{\tau }(1)$
factor there is no term involving $\phi _{3}$. As in eq(\ref{so3}), the QD
mass scale is set by the first term of eq(\ref{win}). For near
degeneracy, the other terms must be relatively small ($y_{\odot } c^2, y_{@} b^2 \ll
y_{0}$, still suppressing the messenger scale).

The charged-lepton masses come from the superpotential 
\begin{equation}
W_{e}=\lambda _{e}(L^{i}\phi _{123}^{i})e^{c}H_{d}+\lambda _{\mu }(L^{i}\phi
_{23}^{i})\mu ^{c}H_{d}+\lambda _{\tau }(L^{i}\phi _{3}^{i})\tau ^{c}H_{d}.
\label{lepton}
\end{equation}%
The $m_{\mu }/m_{\tau }$ ratio is given by $\lambda _{\mu }\langle \phi
_{23}^{i}\rangle /\lambda _{d}\langle \phi _{3}^{i}\rangle$. Using this the
mixing between the second and third families of charged leptons is small of $%
O(m_{\mu }/m_{\tau })$. Similarly one may see that the mixing between the
first and second families is of $O(m_{e}/m_{\mu })$ and that between the
first and third families is of $O(m_{e}/m_{\tau })$, both very small. Ignoring the small corrections from the charged-lepton sector, the
light neutrino mass eigenstates are proportional to the combinations $\phi
_{123}^{i}L^{i} H_u$ and $\phi _{23}^{i}L^{i} H_u$ \footnote{In finding the mass eigenstates with a complex Majorana mass matrix, one needs to be careful to diagonalize $M_\nu M_\nu^\dag$ and not just $M_\nu$.  Because $M_\nu$ is symmetric, it can also be diagonalized by an orthogonal transformation $O M_\nu O^T$.  In general $O \neq U_\nu$ and the square of the eigenvalues of $M_\nu$ are not the same as those of $M_\nu M_\nu^\dag$  \cite{Doi:1980yb}.}.
From eq(\ref{eq:P123 vev}) we
see that these are given by%
\begin{eqnarray}
\nu _{@} &=&\frac{1}{\sqrt{2}}\left( \nu _{\mu }-\nu _{\tau }\right)
\label{neutrino eigenstates} \\
\nu _{\odot } &=&\frac{1}{\sqrt{3}}\left( \nu _{e}+\nu _{\mu }+\nu _{\tau
}\right)  \nonumber
\end{eqnarray}%
where $\nu _{e,\mu ,\tau }$ are the components of $L^{e,\mu ,\tau }$
respectively (selected by the VEV of $H_u$). Ignoring the small charged-lepton mixings discussed above, $\nu
_{e,\mu ,\tau }$ can be identified with the current eigenstates. If $b$ and $c$ are real and positive, and $m_{\odot }=y_{\odot }c^{2}v^{2}<m_{@}=y_{@}b^{2}v^{2}$, one
can see from eq(\ref{win}) and eq(\ref{neutrino eigenstates}) that we obtain the normal
hierarchy, in which $\nu _{@}$ may be identified with the atmospheric
neutrino with bi-maximal mixing while $\nu _{\odot }$ may be identified with
the solar neutrino with tri-maximal mixing. The normal hierarchy persists for a range of complex $b$ and $c$ values in the neighbourhood of the real solution. An inverted hierarchy is possible and viable if $b$, $c$ are approximately imaginary and real, respectively.

Although here we are working at the effective Lagrangian level, we
already noted that $(L^{i}L^{i})HH$ naturally arises from the $SO(3)$
invariant Type II see-saw mechanism. The other two neutrino mass terms can
arise from Type I see-saw through exchange of appropriate heavy right-handed
Majorana neutrinos, in a manner similarly to that discussed for a $SU(3)$
based model in \cite{deMedeirosVarzielas:2005ax}. Being of different origin
it can readily happen that the common mass, $m_{0}=y_{0}v^{2}$ is much
larger than $m_{@}$ and $m_{\odot }$.

\section{Discrete non-Abelian symmetry and vacuum alignment}

We turn now to a discussion as to how the pattern of VEVs displayed in eq(%
\ref{eq:P123 vev}) is dynamically generated. This can be achieved
relatively simply if the underlying family symmetry is a discrete
non-Abelian subgroup of $SO(3)$ (and $SU(3)$). A very simple example is
given by $A_4 \equiv \Delta(12)$, belonging to the $\Delta (3 n^2)$ family of groups \cite{Luhn:2007uq}.
The $\Delta(12)$ invariant terms in the potential are those invariant under the group elements of the semi-direct product $Z_3 \ltimes Z_2$ (which generate the group $\Delta(12)$). The action of these group elements on a triplet representation $\phi^{i=1,2,3}$ is shown in Table \ref{Table2}.

\begin{table}[tbp]
\begin{center}
\begin{tabular}{|l|l|l|}
\hline
& $\mathbf{Z}_{3}$ & $\mathbf{Z}_{2}$ \\ \hline
$\phi ^{1}$ & $\phi ^{2}$ & $\phi ^{1}$ \\ 
$\phi ^{2}$ & $\phi ^{3}$ & $-\phi ^{2}$ \\ 
$\phi ^{3}$ & $\phi ^{1}$ & $-\phi ^{3}$ \\ \hline
\end{tabular}%
\caption{Action of the group factors $Z_3$ and $Z_2$ on the triplet
representation $\phi^i$.}\label{Table2}%
\end{center}
\end{table}%
Since $\Delta(12)$ is a subgroup of $SO(3)$, all $SO(3)$ invariants
are allowed by the discrete subgroup. Thus the terms of eq(\ref{win}) and eq(\ref{lepton}) are allowed. The discrete subgroup allows additional terms, but
these are all higher dimensional and consequently small provided the VEVs of
eq(\ref{eq:P123 vev}) are small relative to the relevant messenger mass.
Thus the lepton mass and mixing structure discussed is a consequence of the
non-Abelian discrete group even though the $SO(3)$ structure used above to
motivate it is only approximate.

Turning now to the question of vacuum alignment, consider the leading terms
in the potential for the triplet familon fields. Because of the $R-$%
symmetry, in the absence of the $X-$field, there are no $F-$terms involving
just the familon fields coming from the superpotential. The leading $D-$%
terms consistent with symmetries of Table \ref{ta:SO3} are 
\begin{equation}
V\left( \phi \right) =\alpha m^{2}\sum_{i}\left\vert \phi ^{i}\right\vert
^{2}+\beta m^{2}\left\vert \sum_{i}\left\vert \phi ^{i}\right\vert
^{2}\right\vert ^{2}+\gamma m^{2}\sum_{i}\left\vert \phi ^{i}\right\vert
^{4}+\delta m^{2} \left\vert \sum_{i} (\phi^{i})^{2} \right\vert ^{2}
\label{potential}
\end{equation}%
Here the quadratic term is driven by supersymmetry breaking and $m$ is the
gravitino mass. The coefficient includes radiative corrections which can
drive it negative at some scale $\Lambda$, triggering a VEV for $\phi$.
The remaining terms can arise through radiative corrections and also only arise
if supersymmetry is broken - hence the factor of $m^{2}$ on every term. The second term is
generated at one-loop order if the superpotential contains a term of the
form $\xi Y\sum_{i}\phi ^{i}\chi ^{i}$ where $\chi ^{i}$ and $Y$ are ($
Z_{3R}=1)$ massive chiral superfields which we take for presentational
simplicity to have mass $M$. These two terms are invariant under the larger
group $SU(3)$ and, if $\alpha$ is negative, generate a VEV of the form $\langle \phi \rangle
=(r,s,t)$ where $r^{2}+s^{2}+t^{2}=x^{2}$, with $x^2$ a constant of $%
O(\Lambda ^{2})$. The third term, consistent with the non-Abelian family
group, breaks $SU(3)$ and $SO(3)$. It will be generated if the underlying theory contains a superpotential term of the form
$((\phi^1)^2 + \omega^2 (\phi^2)^2 + \omega (\phi^3)^2) Z$,
where $\omega$ is the cube root of unity ($\omega^3 =1$) and $Z$ is in a singlet representation of $\Delta(12)$ (one of the three irreducible singlet representations and distinct from the representation of $Y$) \footnote{One may readily check that it is easy to assign charges under the group $G$
to the new fields, $\chi^{i}$, $Y$ and $Z$ 
to allow these couplings.}. This coupling is
invariant under the discrete group but not under $SU(3)$ or $SO(3)$. The
resulting third term of eq(\ref{potential}) splits the vacuum degeneracy.
For negative $\alpha$, the minimum for $\gamma$ positive has $\left\vert
\langle \phi ^{i} \rangle \right\vert =x(1,1,1)/\sqrt{3}$ while for $\gamma $ negative $%
\left\vert \langle \phi ^{i} \rangle \right\vert =x(0,0,1)$. Finally the fourth term also
results from a one-loop radiative correction due to the $\xi Y\sum_{i}\phi
^{i}\chi ^{i}$ interaction. It is $SO(3)$ but not $SU(3)$ invariant and
constrains the phases of the familon fields. For $\delta $ negative and $%
\gamma $ positive the minimum has $\langle \phi ^{i} \rangle = x(1,1,1)/\sqrt{3}$ where $x$
can be complex. This provides a mechanism to generate the vacuum alignment
of $\phi _{3}$ and $\phi _{123}$ as each will have a potential of the form
in eq(\ref{potential}) - as we are considering more than one familon, we label the coefficients with the familon's subscript to identify which term they correspond to. The structure of eq(\ref{eq:P123 vev}) results if $%
\gamma _{3}$ is positive and $\gamma _{123}$, $\delta _{123}$ are negative.

Finally what about $\phi _{23}?$ Its VEV of the form in eq(\ref{eq:P123 vev}%
) readily results once one includes the effect of the $X$ field of Table \ref%
{ta:SO3} because the symmetries allow a term in the superpotential
proportional to $X (\phi _{23} \phi _{123})$. This leads to a positive semi-definite term in the
potential proportional to $\left\vert \phi _{23}\phi _{123}\right\vert ^{2}$.
There remains the need to align $\phi_{23}$ and $\phi_{3}$. This is readily done if radiative corrections generate a term
$m^2 \left\vert \phi^{\dagger}_{23} \phi_{3} \right\vert ^2$ with a negative coefficient, thus a VEV will develop for $\phi _{23}$ in the
direction given by eq(\ref{eq:P123 vev}) \footnote{Note that in eq(\ref{eq:P123 vev}) we have used the freedom to define the directions such that $\langle \phi_{3}^{1,2} \rangle = 0$, $ \langle \phi_{23}^{1} \rangle = 0$.}. One may readily check that the higher dimension terms allowed by the symmetry which involve the $X$ field
always involve an odd factor of $\phi _{23}\phi _{123}$ and do not disturb
the vacuum alignment mechanism discussed above.

\section{Neutrinoless double-beta decay}

\begin{figure}[tbp]
\centerline{\includegraphics[width=3in]{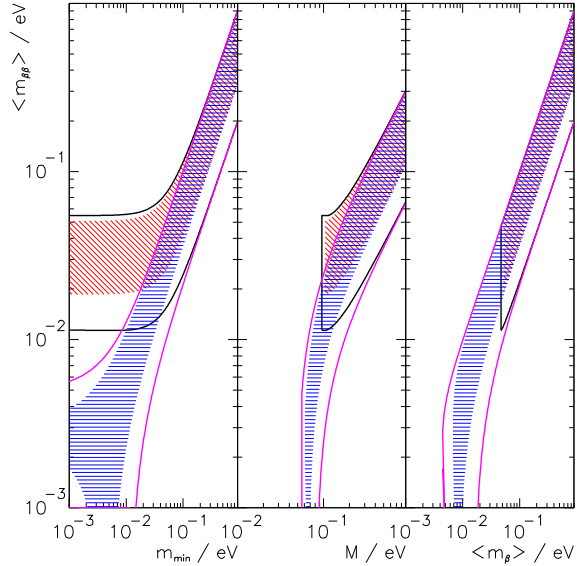}}
\caption{Neutrinoless double-beta decay $m_{\beta \beta}$ plots, from \cite{PDBook2008}. From left to right, $m_{min}$ is the absolute value of the lightest neutrino mass, $M$ is the sum of the light neutrino masses, and $\langle m_{\beta} \rangle$ is the average mass determined from low energy beta decays. The shaded areas has width due to the unknown Majorana phases and the areas enclosed by solid lines take into account the errors of oscillation data. The two sets of solid lines correspond to the normal and inverted hierarchies.}
\label{doublebeta}
\end{figure}

The implication for neutrinoless double-beta decay in this model is
unambiguous because the relative phases of the familon fields are
determined. The amplitude for neutrinoless double-beta decay is proportional
to the magnitude of $\sum m_{\nu _{i}}U_{ei}^{2}\equiv m_{\beta \beta }$ and this is what is measured.
For TBM mixing $U_{e\tau }$ vanishes. The relative phase between
the remaining two terms is given by
$Arg[m_{0}+e^{2i p_{123}}m_{\odot }] - Arg[m_{0}]$ where $p_{123}=Arg[y_{\odot }\phi _{123} \phi_{123} /y_{0}]$.
As $m_{\odot} < m_{0}$ the relative phase remains small.
This corresponds to the
upper branches of Figure \ref{doublebeta} in the QD region.
Complex phases in the VEVs induce other CP violations through the charged-lepton sector that do not significantly affect $m_{\beta \beta }$.

\section{Conclusion}

Attempts to explain the structure of fermion masses and mixings often rely
on structure at a high (Grand Unified?) scale, $M_{F},$ to generate the
observed pattern. One possibility, consistent with neutrino oscillation, is
that neutrinos are nearly degenerate. However, due to enhanced radiative
corrections in this case, the observation of near TBM mixing is difficult to
reconcile with such a high scale mechanism. To keep the deviations from TBM
mixing within experimental limits it is necessary to limit the scale at
which TBM mixing is generated. We have determined this scale for the MSSM
and found significant bounds on $M_{F}$. For example, for degenerate ($m_{\nu
_{i}}>0.1$ eV), normal hierarchy neutrinos and $\tan \beta >6$ (or $\tan \beta >8$, for inverted hierarchy neutrinos), $M_{F}<10^{10}$ GeV is required. To get
close to the Grand Unified scale with QD neutrinos it is necessary to have very small $\tan
\beta$.

Turning to the origin of the structure, we have constructed a model based on
a discrete non-Abelian family symmetry which gives a QD neutrino spectrum
and near TBM mixing. This relies on a natural mechanism for vacuum alignment
of the familons which break the family symmetry. The mechanism predicts that neutrinoless double-beta decay
should be maximal. Although we only constructed the low-energy effective
theory, it fits very well with a see-saw mechanism in which the degenerate
mass comes from a Type II see-saw while the small departures from degeneracy
are driven by a Type I see-saw. 

\section*{Acknowledgments}

IdMV would like to thank Oxford University for their warm hospitality and
support during a short stay when this project was initiated. The work of
IdMV was supported by FCT under the grant SFRH/BPD/35919/2007.
The work of GGR was partially supported by the EC Network 
6th Framework Programme Research and Training Network ``Quest for Unification''
(MRTN-CT-2004-503369) and by the EU FP6 Marie Curie Research and Training Network
``UniverseNet'' (MPRN-CT-2006-035863).
MS acknowledges support from the United States Air Force Institute of
Technology. This work was partly supported by the Science and Technology
Facilities Council of the United Kingdom. The views expressed in this paper
are those of the authors and do not reflect the official policy or position
of the United States Air Force, Department of Defense, or the US Government.


\bibliographystyle{./jhep}
\bibliography{./OxfordResearchReferences}

\end{document}